\documentclass{PoS}

\title{Parsec Scale Properties of Brightest Cluster Galaxies.}

\ShortTitle{Parsec Scale Properties of Brightest Cluster Galaxies.}

\author{\speaker{Elisabetta Liuzzo} $^,$$^1$$^,$$^2$,  Gabriele Giovannini$^1$$^,$$^2$ and Marcello Giroletti$^1$\\
\llap{$^1$}Istituto di Radioastronomia, INAF, via Gobetti 101, 40129
Bologna.\\ 
\llap{$^{2}$}Dipartimento di Astronomia, Universita' di Bologna,via 
Ranzani 1, 40127 Bologna, Italy\\
        E-mail: \email{liuzzo@ira.inaf.it}, \email{ggiovann@ira.inaf.it}, \email{giroletti@ira.inaf.it}}

\abstract{We present new VLBI observations of a complete sample of Brigthest Cluster Galaxies (BCGs) in nearby Abell Clusters. These data show a possible difference between BCGs in cool core clusters (two-sided parsec scale jets) and in non cool core clusters (one-sided parsec scale jet). We suggest that this difference could be due to the jet interaction with the surrounding medium. More data are necessary to discuss whether parsec-scale properties of BCGs are influenced by their peculiar morphology and position at the center of rich galaxy clusters. }

\FullConference{The 9th European VLBI Network Symposium on The role of VLBI in the Golden Age for Radio Astronomy and EVN Users Meeting\\
		 September 23-26, 2008\\
		 Bologna, Italy}

\begin{document}

\section{Introduction}
Brightest Cluster Galaxies (BCGs) are a unique class of objects [11]. These galaxies are the most luminous and massive galaxies in 
the Universe. Most of them are cD galaxies with extended envelopes of excess 
light, but they can be also giant E and D galaxies. The optical morphology often shows evidence of past or recent galaxy mergers (e.g. 
multiple nuclei).  
Moreover, they tend to lie very close to the peak of the cluster X-ray emission 
and in the velocity space they sit near the cluster rest frame. All these 
properties indicate that they could have a quite unusual formation history 
compared to other E galaxies.

In the radio band, BCGs are more likely to be radio-loud than other galaxies of
the same mass [1].  Some BCGs have a standard tailed structure,
either extended on the kpc scale (e.g. 3C465 in A2634), or with a small size
(e.g. NGC4874 in Coma cluster).  In other cases, they show a diffuse and
amorphous radio morphology that is rare in the general radio population, but
very often present in BCGs in cooling core clusters of galaxies. The presence of
X-ray cavities in the emitting gas coincident with the presence of radio lobes
can also be the proof of the interplay between the radio activity of BCGs and
the arrest or slow down of the cooling process at the cluster centers [2,3]. This scenario is in agreement with the recent result that in every/most of cooling core clusters is present an active radio BCG [5].

However, a few points need a deeper study: not all BCGs are strong radio
sources and a restarting activity with a not too long duty cycle is necessary
to justify the slow down of cool~ing processes.  Moreover, it is not clear if radio properties of
BCGs in cooling flow clusters are systematically different from those of BCGs
in merger clusters; we note that in cooling clusters the kpc scale radio morphology
of BCGs is diffuse and relaxed and it is often surrounded by an extended low-brightness radio emission that takes the form of mini-halo [7]. However, extended $`$normal' sources have been also found (e.g. Hydra A, [9]). In merging clusters the most common morphology is a Wide Angle Tail 
source but point-like as well as core-halo sources are also present.  

On the parsec scale, BCGs are not yet well studied as a class of sources.  Only
a few of them have been observed, being well known radio galaxies.  In some
cases, they look like normal FRI radio galaxies with relativistic collimated
jets. Jets are often one-sided because of Doppler boosting effects (e.g.  3C465
in A2634 and 0836+29 in A690 [10]), although there are also
cases where two-sided symmetric jets are present in VLBI images, and the
presence of highly relativistic jets is not certain (e.g. 3C338 in A2199
[4], and Hydra A in A780 [9]).

\section {Observations and Results.}
To study parsec scale properties of BCGs we selected a complete sample of BCGs in nearby Abell clusters with the following constraints: 1) Distance Class $\leq$2, and  2) Declination $> 0^\circ$.
\begin{table}
\caption{{\bf Results for the expanded sample}: in the first column, we report the Abell cluster hosting the BCGs of our sample, in column 2, we put Y if the Abell cluster shows a cool core, N if it doesn't, SCF for small cooling flow and MCF is for middle cooling flow, in column 3 there are the names of BCGs. Column 4 refers to notes where (1) indicates clusters where we observed the BCG with VLBA observations at 6 cm, (2) means that the BCG is a well known radio galaxy with published VLBI data, (3) is for well known BCG outside our complete sample with mas observations. Column 5  is for the large scale morphology of the BCG: we use WAT for Wide Angle Tail radiosource, HT for Head Tail radiosource, MSO for medium symmetric source. In the last column, we mark the parsec scale structure: one sided, two sided or n.d. for the non detections. }
\begin{center}
\footnotesize
\label{tab:id}
\tabcolsep2mm
\begin{tabular}{cccccc}
\hline
\hline
Abell Cluster&cool core&BCG&notes&Large scale&VLBI\\
\hline 
A400&N&3C75A&1&WAT&one sided\\
&N&3C75B&1&WAT&one sided\\
A407&N&UGC2489&1&Tail rs&one sided\\
A539&N&UGC3274&1&radio quiet& n.d.\\
A569&N&NGC2329&1&WAT&one sided\\
A576&N&CGCG261-059&1&Tail rs&one sided\\
A690&N&B2 0836+29 II  &3&WAT&one sided\\
A779&N&NGC2832&1&radio quiet&n.d.\\
A1185&N&NGC3550&1&radio quiet&n.d.\\
A1213&N&4C29.41&1& FRI&one sided\\
A1228&N&IC2738&1&radio quiet&n.d.\\
A1314&N&IC708&1&WAT&one sided\\
&N&IC712&1&small WAT&n.d.\\
A1367&N&NGC3842&1&small WAT&n.d.\\
&N&3C264&2&HT&one sided\\
A1656&N&NGC4874&1&small WAT&one sided\\
A2147&N&UGC10143&1&small WAT& n.d.\\
A2151&N&NGC6041&1&small WAT&core\\
&N&NGC6047&1&compact core+symmetric jets&n.d.\\
A2162&N&NGC6086&1&FRI, relic source &n.d\\
A2197&N&NGC6173&1&point source&one sided\\
A2634&N&3C465&2&WAT&one sided\\
A2666&N&NGC7768&1&Tail rs&one sided\\
\\
A262&Y&NGC708&1&double-no core,jets &core\\
A347&SCF&NGC910&1&radio quiet&n.d.\\
A426&Y&3C84&2&Compact core+Halo& two sided\\
A780&Y&Hydra A&3&double&two sided\\
A1795&Y&4C26.42&3&double&two sided\\
A2052&Y&3C317&3&bright core+halo (FRI)&two sided\\
A2152& MCF&UGC10187&1&Tail rs&n.d.\\
A2199&Y&3C338&2&double restarted&two sided\\
A2390&Y&B2151+174 &3&MSO&two sided\\
A2597&Y&PKS 2322-123 &3&asymmetric radiosource (FRI)& two sided\\
\hline
\hline
\end{tabular}
\end{center}
\end{table}
 All clusters have been included without constraints on cluster conditions (e.g cooling) and no 
selection is present on the BCG radio power. In the complete sample, we have 27 BCGs, including cases like A400
(double BCG, 3C75, Fig. 1) and clusters with a clear double structure (e.g. A1314) where we observed the BCG of both substructures. 



\begin{figure}
\centering
\includegraphics[width=1\textwidth]{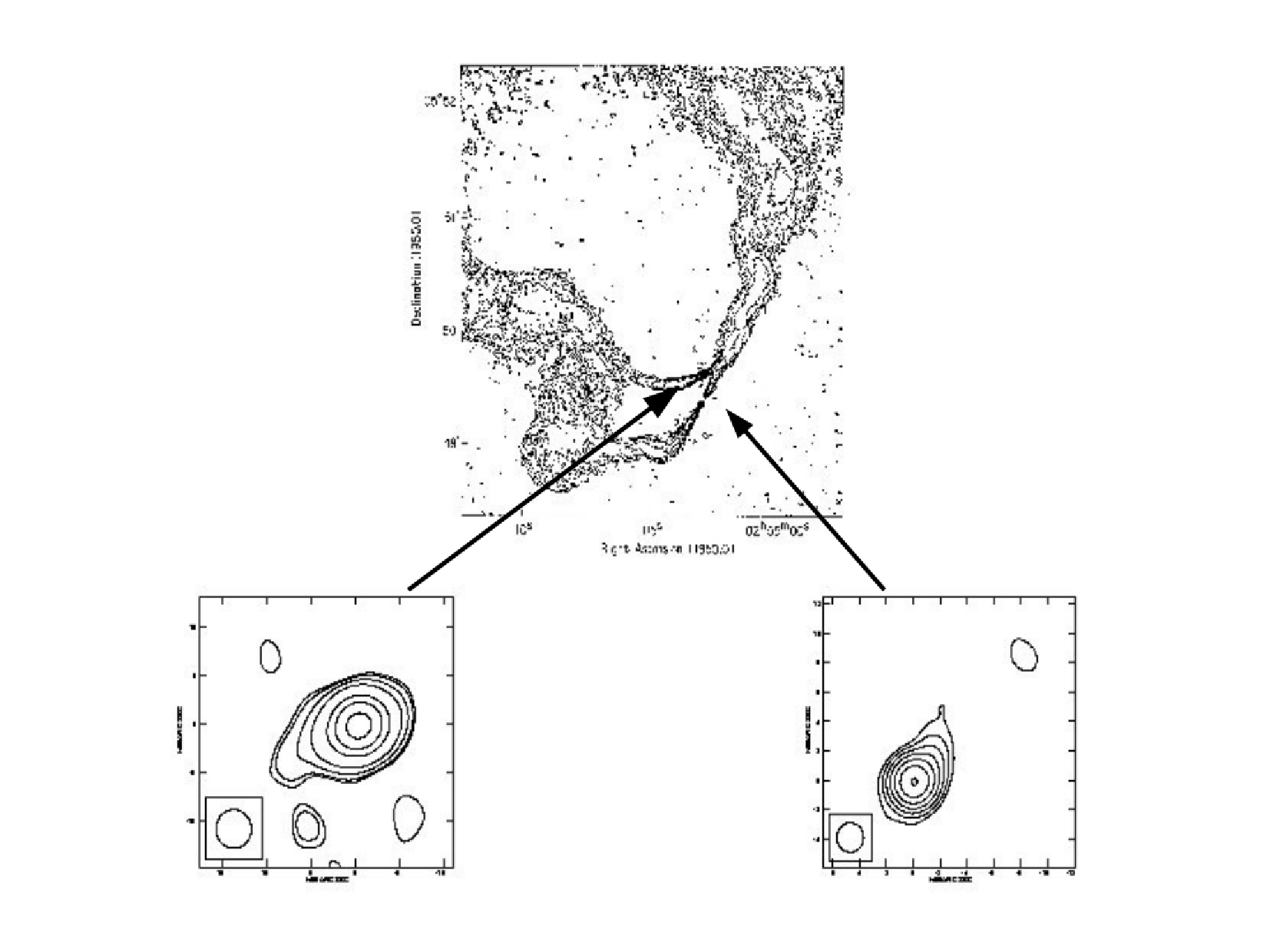}
\caption{In the center: VLA image at 6 cm of 3C 75A and 3C 75B [5], the double BCG in the non cool core cluster A400. Contour intervals are (-1, 1, 2, 4, 8, 16, 32, 64, 128, 256, 512) $\times$ 0.1 mJy/beam. The HPBW is 1.4 $\times$ 1.4 arcsec with P.A. = 0$^{\circ}$. On the right: one-sided VLBI image [8] of 3C75A. Contours levels are 0.3, 0.6, 1.2, 2.4, 4.8, 9.6, 19.2 mJy per clean beam and the HPBW is 2 $\times$ 2 mas with P.A. = 0$^{\circ}$. On the left: one-sided VLBI image [8] of 3C75B. Contours levels are -0.3, 0.3, 0.4, 0.8, 1.8, 7, 14, 28 mJy/beam and the HPBW is 4 $\times$ 4 mas with P.A. = 0 $^{\circ}$. }
\end{figure}

New observations were obtained at the VLBA at 6 cm in phase referencing mode. Each source was observed for about 3 hrs to assure a good uv-coverage and a low noise level. The resolution of the final maps is typically 3 $\times$ 1.8 mas and the noise level is $\sim$ 0.1 mJy/beam. The detection rate in the complete sample (literature and new data) is $\sim$ 60$\%$. Obtained results are shown in Tab.1 together with literature data (expanded sample, see next). A comparison between BCGs in cooling and non-cooling clusters suggests a difference in the properties of the parsec scale structures, but numbers are too small to properly discuss it.

\begin{table}[htp]
\caption{{\bf BCG counts in the complete (nearby) sample and expanded one}.  We
  report the number of BCG according to the cluster morphology and pc scale
  morphology. Note that most of undetected sources in VLBA observations are 
in BCG that
  are radio quiet (or faint) also in VLA observations.}
\begin{center}
\label{tab:id}
\tabcolsep2mm
\begin{tabular}{ccccccc}
\hline
\hline
Sample & Cluster  & Number & two-sided & one-sided & point & N.D. \\
       & morphology &        &           &           &       &      \\
\hline 
Complete & cool core    &  5     &  2 (40$\%$)       &   --      & 1     & 2    \\
         & non cool core    &  22    & --        &  12 (55$\%$)      & 1     & 9    \\
\hline
Expanded & cool core   &  10    & 7 (70$\%$)   & --  & 1  & 2  \\
         & non cool core  &  23    & --  & 13 (56$\%$) & 1  & 9  \\
\hline
\hline
\end{tabular}
\end{center}
\end{table}

To improve our statistics, we
performed a search in the literature and archive data looking for VLBI
data of BCGs in Abell clusters with DC $>$2. We added to our complete sample the following clusters: A690, A780,
A1795, A2052, A2390, and A2597 (expanded sample). Results are presented in Table 1. In the expanded sample, we find  a remarkable  dominance of two-sided sources in relaxed clusters (70\%), and of one-sided  (56\%) or non-detected (39\%) sources in merging systems, as shown in
Tab.\ref{tab:id}.

The difference between relaxed and non relaxed clusters is evident. The presence of one-sided jets could suggest relativistic Doppler boosting. In this hypothesis two-sided jet structures would represent sub-relativistic velocities. However the jet properties can also be related
to different properties in jet formation or to the interaction with a different ISM. In BCG at the center of cooling clusters the gas density in the ISM region is expected to be higher. Therefore we can assume a strong interaction of the jet at parsec resolution with the environment.
\begin{figure}[htp]
\centering
\includegraphics[width=0.4\textwidth]{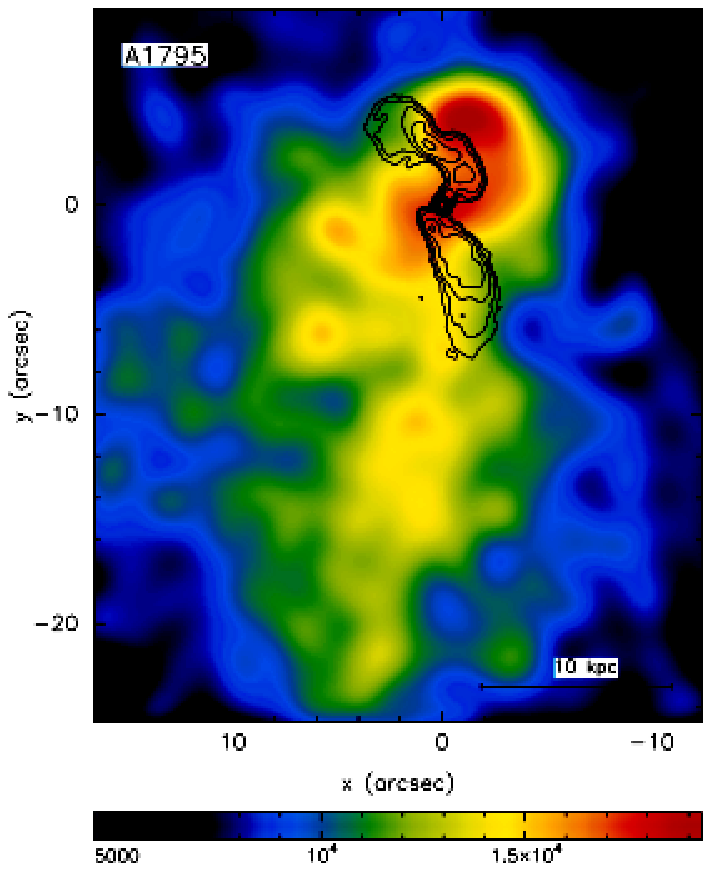}
\hfill
\includegraphics[width=0.4\textwidth]{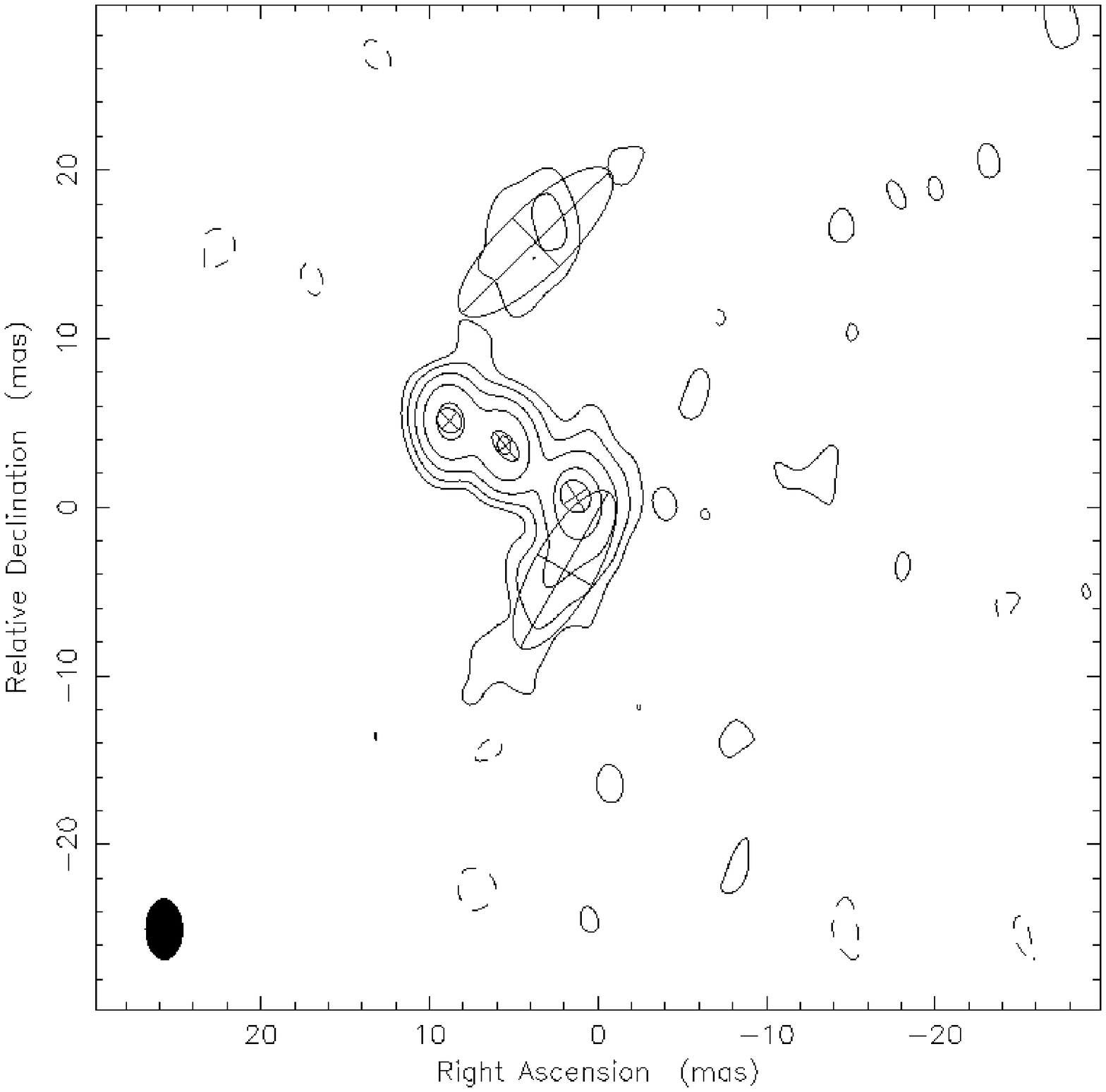}
\caption{On the left: an overlay [6] of the 3.6 cm radio emission of 4C 26.42 (BCG of Abell 1795) at 0.37-arcsec resolution
on the Chandra X-ray image convolved with a 2-arcsec Gaussian. The radio contours
start at 0.08 mJy/beam and increase by factors of 2 to just below the peak
in the image of 33 mJy/beam. On the right: two-sided VLBI image at 5 GHz with components given by Modelfit. Contours of VLBI image are -0.4, 0.4, 0.8, 1.6, 3.2, 6.4 mJy/beam and the HPBW is 3.6 $\times$ 2.2 mas with P.A. = 0.543 $^{\circ}$.}
\end{figure}
This difference of radio morphology between BCGs in cool core and non cool core clusters could be also related to 
a different availability of gas in the nuclear regions. This suggestion is supported by literature data on BCGs of more distant clusters as e.g. 4C26.42 (Fig.2), the BCG in the cool core cluster A1795, where VLBI images show a distorted symmetric structure [8], and Hydra A (the BCG of the cool core cluster A780), where Taylor (1996) [9] suggested that the emission from the symmetric parsec scale jets is more dependent on interactions with the surrounding material than on Doppler boosting. 

On the other hand, the one-sided structure in non cool core clusters might due to Doppler boosting effects in relativistic, intrinsically symmetric jets. This is the case for example of 3C 75 A and B (Fig.1), 3C465 [10] and more. 

We conclude that the possibile dichotomy between BCGs in cool core and non cool core clusters could be due not to intrinsic jet differences but to different ISM conditions. We plan to observe a larger sample of BGCs in cooling and relaxed clusters with VLBA to improve our statistic.

\section{Acknowledgments.}
We thank the organizers of a very interesting meeting. The National Radio Astronomy Observatory is operated by Associeted Universities, Inc., under cooperative agreement with the National Science Foundation.

\end{document}